# Statistical witch-hunts:
Science, justice & the p-value crisis

Spencer Wheatley and Didier Sornette
Chair of Entrepreneurial Risks, D-MTEC, ETH Zürich

**Summary**

We provide accessible insight into the current 'replication crisis' in statistical science, by revisiting the metaphor of court trial as hypothesis test. Inter alia, we define and diagnose harmful *statistical witch-hunting* both in justice and science.

Justice is, in principle, a statistical science, and trial as hypothesis test is classic pedagogy. We modernize the link here, noting that equivalent errors, and challenges exist in both justice and applied statistics. We introduce a severe but applicable metaphor for irresponsible statistical inference: *statistical witch-hunting* – also applying to justice, elevating risk of harming innocents. It follows that justice would benefit from further assimilating tools and principles of statistics. By riddle: 'W*hy do witch-hunters find fewer witches when they bring a statistician with them?*'

The same (and more) can be said of applied statistical science, with its replication crisis. Drawing from eminent theoretical and applied statistician Frank Hampel's "*Is statistics too difficult?*"[1]: Statistics and its application has led to great advances – e.g., in technology, scientific knowledge, and decision-making – forming "a necessary part of our civilization". As with all technology, "statistics also has another, darker side", with misuse and misunderstanding long being a widespread problem: inter alia, corrupting 'scientific knowledge' with spurious results.

Good statistical science[2], requiring education and experience, is a highly intelligent and creative process, for which demand outstrips supply – in academia and industry, which are big, and increasingly data-intensive. Most applied statistics is done by researchers within their own disciplines. And, by volume, most statistics is not even done by humans, but by data-mining machines. Broad use and automation have led to accessible software for routine methods. However, routine methods alone are often inadequate for real-world problems[3]. It has become easy to obtain results by simply pushing a button, with virtually no 'scientific rigor' necessary.

Under widespread adoption, statistical science struggles to maintain rigor. Serious improvements in education, methodology, review, and so on, have been proposed. But ongoing claims to ban p-values and statistical significance seem radical. Indeed: 1) where p-values are not used properly, there will be more serious underlying problems ('tip of iceberg'), and 2) statistical decisions need to be taken: i.e., without them there can be no court verdict, the doctor cannot tell the patient if they are sick, and the policy-maker cannot decide if climate change is worthy of action. Those pushing ban p-values should consider the risks, especially to practitioners and decision-makers. Efforts should be focused on the (hard) root problems causing low quality statistical science – not convenient scapegoats. This is key to support the continued productive evolution and exploitation of statistical science and technology. For instance, we echo the call of John Tukey, made more than 50 years ago, for a more scientific orientation of academic statistics – in particular towards a high quality 'data science' as outlined by David Donoho[4].

---

[1] Hampel, Frank. "Is statistics too difficult?." *Canadian Journal of Statistics* 26.3 (1998): 497-513.
[2] Requiring, not least, a full understanding of the applicability of the stochastic model/technology, incl. how wrong the model approximation of reality is, and the consequences of that error.
[3] As observed by Hampel in his statistical consulting experience.
[4] Donoho, David. "50 years of data science." Journal of Computational and Graphical Statistics 26.4 (2017): 745-766.

## 1. Prologue: On witch-hunting & human nature

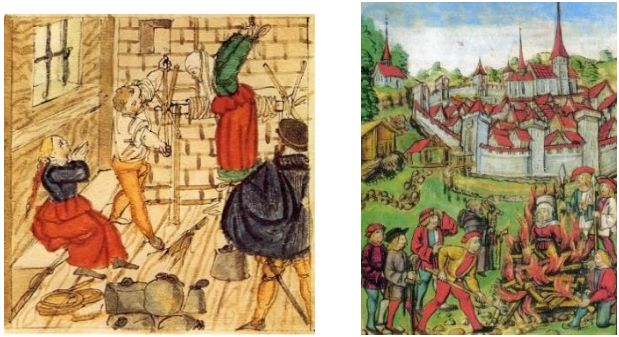

Figure 1. Left: Torturing of a witch in Baden, Switzerland (1585), by Johann Jakob Wick. Right: The burning of a woman in Willisau, Switzerland, by Diepold Schilling (1447). Curiously, in the middle ages, the Swiss executed the most "witches" per capita in Europe[5].

There will always be a desire to blame scapegoats for disasters or other misfortunes lacking a *convenient* explanation. In the 'old days', abnormal behaviour, or personal motives of an accuser could lead to someone being accused of satanic witchcraft, leading to a *trial/test*. The witch-hunters had a battery of tests and tortures available, designed to achieve the desired outcome[6]: E.g., *1) Trial by water*: If the accused floated when thrown in a river then they were a witch! If they were 'lucky enough' to sink, then they were innocent. 2) *Prayer test*: If unable to calmly and fluently recite scripture, they were a witch! 3) *Marks and pricks:* Any physical imperfections, such as moles, could be taken as evidence. It was assumed that these marks would be insensitive to pain and wouldn't bleed. Professional "prickers" ran tests – the more mercenary ones using dull needles to ensure 'false positives'. Despite progress, human nature remains, and we should strive to avoid witch-huntings' more modern manifestations.

## 2. Justice as statistical science: modern parallels

Striving for real examples to illuminate abstract statistical concepts[1] we start with the fundamental but often misused NHST[7] (*null hypothesis significance testing*), relating it to a court trial, and then exploit a more general analogy. The *world of the courtroom* and the *world of statistical science* are materially linked by but a few threads. However they share a fundamental nature: they both construct hypotheses and test them with evidence. 'NHST as court trial' is a classic pedagogical example: a test where a guilty verdict is issued if the evidence is sufficiently discordant with the presumption of innocence – further developed in Table 1.

| Hypothesis testing | Courtroom |
| --- | --- |
| Null hypothesis. | Presumption of innocence. |
| Reject the null hypothesis. | Issue guilty verdict. |
| Test level, e.g., p-value<0.05. | Standard of proof/'beyond a reasonable doubt'. |
| Test statistic. | Summary of evidence. |
| Statistical test. | Jury deliberation and decision. |
| Type 1 error / type 2 error. | False conviction / false acquittal. |
| Prior belief / distribution. | E.g., biased jury. |
| Invalid test. | Mistrial. |

Table 1. Interpretation of the classical analogy[8] of hypothesis testing and a criminal trial. Neyman and Pearson[9] even considered the analogy when pioneering the theory.

---

[5] I. Eichenberger. "No one tortured witches like the Swiss". Swissinfo.ch (14, Sept, 2009)
[6] E. Andrews. "7 Bizarre Witch Trial Tests". History.com (14, Mar, 2014).
[7] E.L Lehmann, and J.P. Romano. Testing statistical hypotheses. Springer Science & Business Media, 2006.
[8] M.A. Martin, (2003). "'It's like … you know': The Use of Analogies and Heuristics in Teaching Introductory Statistical Methods." The Journal of Statistics Education, 11(2)
[9] J. Neyman, and E.S. Pearson. (1933). "On the Problem of the Most Efficient Tests of Statistical Hypotheses." Philosophical Transactions of the Royal Society A, pp. 289-337.

To emphasize the gravity and necessity of *statistical decisions*: In justice, a classic debate exists over the optimal stringency of the *standard of proof,* relating to the moral imperative of harm reduction. False convictions exist (type 1 error), but have to be balanced with the risk of false acquittals (type 2 error). With no proof required, it would be easy to convict many criminals, but also many innocents. With absolute proof required, most criminals would walk free. *Statistical decision theory* allows to determine the optimal level, being equal to the probability of 'false conviction': e.g., with level 0.05, independently repeating trial of an innocent, one expects false conviction 5% of the time. Legal scholars have lamented the unknowability of the false conviction rate. But, with statistics, it was estimated as 4.3%[10] for 'death row, USA' – a 'cosmic coincidence' that the court has self-organized towards the 5% level – a 'default' in many fields.

In 2004, an American professor of law and justice prefaced his book with a comment on the immature state of statistics in justice: "…we have a coherent sophisticated, effective framework managing errors in statistical inference, but no such framework in the criminal justice system"[11]. However, the commonality between a court trial and 'statistical science' in practice becomes clearer when extending the analogy (see table 2): the familiar courtroom terms make their statistical counterparts immediately understandable. In brief, any criminal trial is far more complicated than a simple hypothesis test, however the same is true of any scientific study.

| Statistical science | Courtroom/Justice |
|---|---|
| Faulty experimental design. | Flawed investigation/evidence discovery. |
| Data manipulation/selection. | Evidence tampering. |
| *Data-snooping*: Selecting test after seeing data. | Selecting the charge after knowing jury/evidence. |
| Selecting the test that confirms your hypothesis. | Jury rigging, or worse: higher level corruption. 'Guilty until proven innocent'. |
| *Witch-hunting*: Multiple testing and data-mining. | Witch-hunting: 'trial until guilty'. |
| Journal editors, review committee. | Judge, appel courts, etc. |
| *Moral hazard*: Pressure in pharmaceutical trials to rush the product to market; political bias in social sciences; cheating for career advancement, etc. | *Moral hazard*: Political bias, commercial interests, money, and so on irrefutably influence the outcome of trials. |
| False discovery published: Spurious publications may be long-lived[12], being both used and cited, as contradictory negative findings are seldom published, and published works seldom retracted. | *False precedent*: Errors accepted into common law may propagate far, causing harmful decisions, due to principle '*stare decisis*' to abide by precedent and not disturb settled matters. |
| Study rejected/retracted (e.g., by journal). | Verdict overturned/appeal. |
| Ongoing *replication crisis*, where published significant results cannot be reproduced; in some fields in the majority of cases.[13] | Scenario of irresponsible mis-use of justice, leading to many innocents being convicted, and loss of confidence in the judicial authority & process. |

Table 2. A modern analogy of applied statistical sciences and the criminal justice process in terms of mis-uses and errors. The final row is a summary issue, to which all the other mis-uses/abuses may contribute.

---

[10] S.R. Gross, et al. "Rate of false conviction of criminal defendants who are sentenced to death." *Proceedings of the National Academy of Sciences* 111.20 (2014): 7230-7235.
[11] B. Forst. "*Errors of justice: Nature, sources and remedies*. Cambridge University Press, 2004.
[12] K. Cor and G. Sood, 2019 find that about 30% of positive citations to retracted articles occur in the year following their retraction ("Propagation of Error: Approving Citations to Problematic Research": http://www.gsood.com/research/papers/error.pdf)
[13] https://en.wikipedia.org/wiki/Replication_crisis ; Harris, Richard. *Rigor mortis: how sloppy science creates worthless cures, crushes hope, and wastes billions*. Basic Books, 2017.

Further, the justice world deals with a number of challenges that pervade statistical science:

- Communicating statistical evidence to non-expert decision-makers (e.g., communication of uncertain risk information to politicians, like uncertain evidence to the jury).
- Quantification of belief (e.g., through subjective probabilities).
- Drawing conclusions from diverse/complex evidence (e.g., with Bayesian networks: graphical models of probabilistic relationships between hypotheses and evidence).[14,15]
- Lack of important data due to sensitivity/confidentiality.
- Handling human cognitive errors (e.g., jurors, investors, or power plant operators).
- Differentiating between statistically significant and practically meaningful results[16].

Indeed, the esoteric field of *forensic statistics* focuses on statistics in the justice system. According to Prof. Colin Aitken: The field is small, with work – e.g., supporting analyses and providing expert witness – done by academics and others on a consulting basis.[17] Despite the clear importance of the field, the small community faces a battle against courts that resist adoption of statistical sciences.[18] Nevertheless, foundational work has been done and is ongoing.[19,20]

### 3. Statistical witchhunts in justice

To borrow a quote from Prof. Norman Fenton, **"*Proper use of probabilistic reasoning has the potential to improve the efficiency, transparency and fairness of the criminal justice system*."**

We now exploit statistical principles to support better justice. To start, in every court trial – and interactions with authority, under the risk of undeserved harm[21] – the judicial system plays 'Russian roulette' with the freedom, livelihood, and reputation of the accused. Indeed one can argue that any time a person is put on trial and found not guilty, they should be compensated – e.g., on a pure risk basis, 0.05 times the 'cost' associated with a false conviction. And, if determined to be based on a false accusation, the malicious accuser should pay (at least) that.

Next, consider the *multiple testing problem*[22] of modern statistics: If a statistical experiment is repeated enough times, eventually (by chance) a false significant result will be found. It is plausible that this 'false-conviction' error underlies many of the scientific studies that cannot be reproduced. E.g., American drug company, Amgen, tried to reproduce the successes of 53 'landmark' cancer papers, but recovered a mere 11% of the results[23]. A prevalent manifestation of multiple testing, especially in finance, is that, e.g., given a big set of unrelated time series, sets of 'significantly correlated' series will be spuriously identified. E.g., a portfolio manager "sifted through a United Nations CD-ROM and discovered that historically, the single best prediction of the Standard & Poor's 500 stock index was butter production in Bangladesh."[24] As a joke, he published the finding. People still earnestly contact him to exploit this miraculous correlation.

---

[14] Y. McDermott, and C.G. Aitken. "Analysis of evidence in international criminal trials using Bayesian Belief Networks." *Law, Probability and Risk* 16.2-3 (2017): 111-129
[15] F. Taroni, C.G.G Aitken, P. Garbolino, & A. Biedermann. (2006). Bayesian networks and probabilistic inference in forensic science (p. 372). Chichester: Wiley.
[16] C.G. Aitken, A. Wilson, and R. Sleeman. "Statistical significance—meaningful or not." *Law, Probability and Risk* (2018)
[17] https://www.statslife.org.uk/careers/profiles/1135-colin-aitken-forensic-statistician
[18] N. Fenton. "Science and law: Improve statistics in court." Nature 479.7371 (2011): 36.
[19] C.G.G Aitken, and F. Taroni. "Fundamentals of statistical evidence—a primer for legal professionals." *The International Journal of Evidence & Proof* 12.3 (2008): 181-207.
[20] D.A. Schum. The evidential foundations of probabilistic reasoning. Northwestern University Press, 2001.
[21] 'Stop and frisk', or police raids provide salient examples.
[22] P.C O'Brien, and T.R. Fleming. "A multiple testing procedure for clinical trials." Biometrics (1979): 549-556.
[23] M. Baker. "Biotech giant publishes failures to confirm high-profile science", Nature 530, 141 (2016).
[24] R. Sullivan, et. al. "Data-Snooping, Technical Trading Rule Performance, and the Bootstrap", The Journal of Finance 54 (5)(1999)

The manifestation of multiple testing in justice is clear: If running multiple trials for the same crime, or against the same person, an appropriately higher burden of proof (e.g., *Bonferroni correction*[25]) should be required in each of the individual trials to avoid the probability of a false conviction from mushrooming. Ignoring these basic statistical principles, a judicial witchhunt could apply "trial until guilty" – a complete abuse of justice that tortures the presumption of innocence until it collapses.

So, if going on a witch-hunt – in the name of science or justice – please bring a *good statistician* along with you. You may be disappointed with the catch, but there is a good chance that less harm will be done.

### 4. The crisis of statistical science: witch-hunters turn on their own wares?

Next, is statistical science – with a dominant share done by 'non-statisticians', or even data-mining machines – in worse shape than justice? 'The field', which is extremely diverse, is in crisis, having a false positive rate of publications many-fold-higher than the 4.3% false conviction rate on US death row. But what are *the authorities* saying? First, looking to the American Statistical Association (ASA): In 2016, the ASA issued guidelines on use of p-values.[26] The basic idea was that one should not rely on p-values alone – i.e., neglecting all other relevant information such as sample size[27], quality of model and study, and size of estimated effect. In effect, this is an injunction for a *minimal level of rigor*, due to an observed deficit in practice. Serious and substantial literature followed.

In March, 2019, the ASA pushes further 'reforms' via editorial, entitled, "*Moving to a World Beyond p < 0.05*". It concluded *"based on our review […] it is time to stop using the term "statistically significant" entirely. Nor should variants […] survive."* Assessing the state of statistical science as "*old, rotting timbers*", they discuss important issues such as best practices, as well as institutional reform. Remedial prescriptions such as "*accept uncertainty*", "*be open, and modest*" and do "*thoughtful science*", again simply discuss basic rigor, although without reference to standard philosophy of science. But, their focus and conclusions target use of p-values: finishing on a revolutionary note, "*Let's move beyond "statistically significant," even if upheaval and disruption are inevitable […] It's worth it […] by breaking free from the bonds of statistical significance, statistics in science and policy will become more significant than ever.*"

Next, from the prestigious journal "Nature"[28]: "*We agree and call for the entire concept of statistical significance to be abandoned.*" This is then apparently contradicted with, "*We are not advocating a ban on P values, confidence intervals or other statistical measures*", and finally "*Our call to retire statistical significance and to use confidence intervals as compatibility intervals is not a panacea. Although it will eliminate many bad practices, it could well introduce new ones.*" The journal followed up with an editorial "*It's time to talk about ditching statistical significance*"[29]. Apparently more than 850 'scientists' have endorsed the article in the online commentary facility.

---

[25] Dunn, Olive Jean. "Multiple comparisons among means." *Journal of the American statistical association* 56.293 (1961)
[26] R.L. Wasserstein, and N.A. Lazar. "The ASA's statement on p-values: context, process, and purpose." The American Statistician 70.2 (2016): 129-133.
[27] E,g., debated between: P. Vermeesch. (2009), Lies, Damned lies and statistics (in geology), EOS 90 (47), 443 (2009) , and D. Sornette and V. Pisarenko, On the Correct Use of Statistical Tests: Reply to "Lies, damned lies and statistics (in Geology)", Eos Transactions AGU, Vol. 92, No. 8, p. 64 (22 February 2011)
[28] V. Amrhein, S. Greenland, and B. McShane, "Scientists rise up against statistical significance". Nature 567, 305-307 (2019)
[29] "It's time to talk about ditching statistical significance", Nature 567, 283 (2019)

Behind the smoke, is there fire? Indeed, and no less than a lack of quality in scientific work. Good science requires rigor across its full lifecycle. Peng and Leek made this point, loud and clear, with "p-values are just the tip of the iceberg"[30]. Indeed and furthermore, studies must be consistently assimilated with the body of literature, and truth elevated. This goes beyond technology, involving system design, philosophy, education and human behavior—there is no panacea to ensure 'good science'. We can be open to paradigm shift[31] in statistical science, but disruptive revolution should be used with great care and only when truly needed. Appropriately, serious remedial actions have been proposed: encouraging replication studies, publication of negative findings, open data, new statistical tools, and so on. From here, 'burning p-values at the stake' looks a bit like reactive scapegoating, whereas focusing on the (hard) root problems will be the way to support the continued productive evolution and exploitation of statistics in science and beyond.

We conclude by considering the experience of the Catholic church -- under which much of western science took place, especially before the printing press[32]. In particular, from Pope Paul VI:

*" …from some fissure the smoke of Satan has entered the temple of God. There is doubt, incertitude, problematic, disquiet, dissatisfaction, confrontation. There is no longer trust of the Church; they trust the first profane prophet who speaks in some journal or some social movement, and they run after him and ask him if he has the formula of true life. And we are not alert to the fact that we are already the owners and masters of the formula of true life. Doubt has entered our consciences, and it entered by windows that should have been open to the light. Science exists to give us truths that do not separate from God, but make us seek him all the more and celebrate him with greater intensity; instead, science gives us criticism and doubt. Scientists are those who more thoughtfully and more painfully exert their minds. But they end up teaching us: "I don't know, we don't know, we cannot know." The school becomes the gymnasium of confusion and sometimes of absurd contradictions. Progress is celebrated, only so that it can then be demolished with revolutions that are more radical and more strange, so as to negate everything that has been achieved, and to come away as primitives after having so exalted the advances of the modern world."* – Pope Paul VI, 1972[33]

---

[30] J.T. Leek, and R.D. Peng. "Statistics: P values are just the tip of the iceberg." Nature News 520.7549 (2015): 612.
[31] T.S. Kuhn. *The structure of scientific revolutions*. University of Chicago press, 2012.
[32] https://en.wikipedia.org/wiki/Catholic_Church_and_science
[33] Pope Paul VI, "Mass on the 9th anniversary of the crowning of His Holiness Paul VI on the Solemnity of the Apostles Peter and Paul", 29 June 1972, https://w2.vatican.va/content/paul-vi/en/homilies/1972.index.html